\def \SAIT #1 #2 {{\em Mem.\ Soc.\ Astron.\ It.\/} {\bf #1}, #2}
\def \MESS #1 #2 {{\em The Messenger\/} {\bf #1}, #2}
\def \ASTRNACH #1 #2 {{\em Astron. Nach.\/} {\bf #1}, #2}
\def \AAP #1 #2 {{\em Astron. Astrophys.\/} {\bf #1}, #2}
\def \AAL #1 #2 {{\em Astron. Astrophys. Lett.\/} {\bf #1}, L#2}
\def \AAR #1 #2 {{\em Astron. Astrophys. Rev.\/} {\bf #1}, #2}
\def \AAS #1 #2 {{\em Astron. Astrophys. Suppl. Ser.\/} {\bf #1}, #2}
\def \AJ #1 #2 {{\em Astron. J.\/} {\bf #1}, #2}
\def \ANNREV #1 #2 {{\em Ann. Rev. Astron. Astrophys.\/} {\bf #1}, #2}
\def \APJ #1 #2 {{\em Astrophys. J.\/} {\bf #1}, #2}
\def \APJL #1 #2 {{\em Astrophys. J. Lett.\/} {\bf #1}, L#2}
\def \APJS #1 #2 {{\em Astrophys. J. Suppl.\/} {\bf #1}, #2}
\def \APSS #1 #2 {{\em Astrophys. Space Sci.\/} {\bf #1}, #2}
\def \ASR #1 #2 {{\em Adv. Space Res.\/} {\bf #1}, #2}
\def \BAIC #1 #2 {{\em Bull. Astron. Inst. Czechosl.\/} {\bf #1}, #2}
\def \JSQRT #1 #2 {{\em J. Quant. Spectrosc. Radiat. Transfer\/} {\bf #1}, #2}
\def \MN #1 #2 {{\em Mon. Not. R. Astr. Soc.\/} {\bf #1}, #2}
\def \MEM #1 #2 {{\em Mem. R. Astr. Soc.\/} {\bf #1}, #2}
\def \PLR #1 #2 {{\em Phys. Lett. Rev.\/} {\bf #1}, #2}
\def \PASJ #1 #2 {{\em Publ. Astron. Soc. Japan\/} {\bf #1}, #2}
\def \PASP #1 #2 {{\em Publ. Astr. Soc. Pacific\/} {\bf #1}, #2}
\def \NAT #1 #2 {{\em Nature\/} {\bf #1}, #2}

\documentstyle{memsait}
\input epsf.sty
\begin{opening}
\title{LOOKING FOR HIGH ENERGY PEAKED BLAZARS} 
\author{L. Costamante$^{1,2}$, G. Ghisellini$^2$, A. Celotti$^3$,
P. Giommi$^4$, P. Padovani$^5$,\\
 G. Tagliaferri$^2$, A. Wolter$^2$, M. Chiaberge$^4$, G. Fossati$^6$, 
E. Pian$^7$, \\
L. Maraschi$^2$, F. Tavecchio$^2$, A. Treves$^8$, }
\institute{$^1$University of Milan, Italy; 
$^2$Osservatorio di Brera, Milan, Italy; $^5$STScI, Baltimore, USA; 
$^3$S.I.S.S.A., Trieste, Italy; $^4${\it Beppo}SAX SDC, Rome, Italy;
$^6$CASS/UCSD, La Jolla, USA; $^7$TESRE, Bologna, Italy; 
 $^8$Univ. dell'Insubria, Como, Italy.}
\date{} 
\end{opening}

\begin{document}

\oddpagefooter{}{}{} 
\evenpagefooter{}{}{} 

\bigskip

\begin{abstract}
Blazars can be classified on the basis of their overall
Spectral Energy Distribution (SED).
BL Lac objects are usually divided in LBL or HBL (Low or High energy peaked
BL Lacs), according to the peak frequency of the synchrotron emission, if 
in the optical or UV--soft-X band respectively. 
FSRQs instead are characterized by synchrotron peaks mainly at 
IR--optical frequencies, similarly to LBLs.
Here we report on recent {\it Beppo}SAX observations which are unveiling
the high energy branch of the range of
synchrotron peak frequencies.
Four new ``extreme" HBLs have been discovered, one of which (1ES 1426+428) 
peaks near or above 100 keV, in a quiescent state.
\end{abstract}

\section{Introduction}
Blazars can be classified also
by the shape of their overall SED, as ``red" or ``blue" objects 
 whether the peak of the synchrotron 
emission (in a $\nu F_{\nu}$ representation) lies in the IR--optical 
 or UV--soft-X band, respectively. 
In the X--ray band this distinction becomes 
quite apparent, since ``red" objects show a flat ($\alpha_x<1$) 
inverse Compton spectrum, while ``blue" objects show a steep ($\alpha_x>1$) 
spectrum, due to the tail of the synchrotron emission.
It was believed that FSRQs showed exclusively ``red" properties,
while BLLacs were found with either ``red" (LBLs, mainly radio selected)
or ``blue" properties (HBLs, mainly X--ray selected).
Recently however, new surveys like the DXRBS (Perlman et al. 1998), the RGB 
(Laurent--Muehleisen et al. 1998) and the REX (Caccianiga et al. 1999)
are revealing a more complete scenario,
thanks to the covering of regions of the parameters space previously 
unexplored (in $F_x$, $F_r$ and broad band 
spectral indices $\alpha_{rx}$, $\alpha_{ro}$ and $\alpha_{ox}$).
Among FSRQs, they have discovered a new type of objects, with broad band spectral 
indices very similar to HBLs, which qualify them as 
``blue quasar" candidates, the missing counterpart in
a possible FSRQ--BL Lac simmetric scenario
(further details and the {\it Beppo}SAX discover of the
first ``blue" FSRQ in Padovani et al. 2000, in preparation).
Among BLLacs, these surveys are showing that they actually form a 
continuous class with 
respect to the peak of the synchrotron emission, which smoothly ranges 
between IR and soft--X frequencies (and up to
the 2--10 keV band for some ``extreme" sources like Mkn 421).
Another interesting question is if there exists a limit
at the blue end of the sequence.
In fact, the
{\it Beppo}SAX observations of Mkn 501 (Pian et al. 1997) 
and 1ES 2344+514 (Giommi et al. 1997) have revealed that, 
at least in a
flaring state, the synchrotron peak can reach 
even higher 
energies, around or above 100 keV.
In these cases the X--ray slope in the 1--10 keV band
is {\it flat} even if the radiation process is synchrotron.
Here we report on the {\it Beppo}SAX observing campaign
performed with the aim of finding  and studying other sources 
as ``extreme"
as Mkn 501 in flaring state.

%
\vspace{0.5cm} 
\centerline{\bf Tab. 1 - LECS+MECS  best fit parameters}
\begin{table}[h]
\begin{tabular}{|l|c|c|c|c|c|c|c|}
\hline
   & & & & & & &
\vspace{-3.3mm} \\
Source  & N$_{\rm H}$ & $\alpha_1$ & E$_{break}$ & $\alpha_2$ & $F_{1keV}$ &
$F_{2-10}$  & $\chi^2_r$/d.o.f. \\
 & $10^{20}$ cm$^{-2}$ & & keV & & $\mu$Jy &  ergs/cm$^{2}$s  & \\
\hline
  & & & & & & &   
\vspace{-3mm} \\
1ES 0120+340 & 
$5.2 \,{\it gal.}$  & $0.82^{-0.96}_{+0.26}$
 & $1.4^{-0.7}_{+1.2}$ & $ 1.32^{-0.08}_{+0.08}$ & $4.5^{-0.6}_{+2.1}$ &
 $1.3\; \times10^{-11}$ & 0.92/93 \\  
  & & & & & & &
 \vspace{-3mm} \\
PKS 0548-322 &
$4.2^{-0.9}_{+1.1}$ & $0.91^{-0.16}_{+0.10}$ &
$4.4^{-2.2}_{+1.8}$ & $1.38^{-0.31}_{+0.59}$ & $5.7^{-0.5}_{+0.5}$ &
$2.3\; \times10^{-11}$ & 0.95/82 \\
  & & & & & & &
 \vspace{-3mm} \\
1ES 1426+428 & 
$1.5^{-0.3}_{+0.4}$ & $0.92^{-0.04}_{+0.04}$ & --- & --- &
$4.6^{-0.2}_{+0.2}$ & $2.0\; \times10^{-11}$ & 1.00/89 \\
  & & & & & & &
  \vspace{-3mm} \\
H 2356-309 &
$1.3 \,{\it gal.} $ & $0.78^{-0.09}_{+0.06}$ &
$1.8^{-0.6}_{+0.6}$ & $1.10^{-0.05}_{+0.05}$ & $6.2^{-0.5}_{+0.5}$ &
$2.5\; \times10^{-11}$ & 0.94/35 \\
%
\hline

\end{tabular}
\end{table}

\section{Extreme BL Lacs}
Our candidates have been selected from the Einstein Slew Survey and the
RASSBSC catalogues.
The selection criteria were based on properties  suggesting a high
$\nu_{peak}$: 
a) very high $F_x / F_{radio}$ ratio ($>3\times10^{-10}$  erg cm$^{-2}$
s$^{-1}$ / Jy,  at [0.1--2.4] keV and 5 GHz, respectively);
b) flat X--ray spectrum (when available), connecting smoothly with
the flux at lower frequencies;
c) appropriate values of $\alpha_{ro}$, $\alpha_{ox}$ and $\alpha_{rx}$
(Padovani \& Giommi 1995). A high X--ray flux ($>10^{-11}$ erg cm$^{-2}$
s$^{-1}$) was also requested, to achieve a good
detection in the PDS instrument. 
In Table 1 the main results for the first 4 sources observed by 
{\it Beppo}SAX are reported. 
All have been detected in the PDS band.
For three of them the spectrum is best fitted  with a convex broken power--law:
this locates the peak of the synchrotron emission 
at 1--4 keV, thus confirming the ``extreme" nature of these sources.
The spectrum of 1ES 1426+428 is  instead well fitted by a single power--law,
with a flat spectral index ($\alpha=0.92$) up to 100 keV, thus 
{\bf constraining the synchrotron peak to lie near or above 100 keV}.
This result was not straightforward, due to the presence
of the quasar GB 1428+422 (Fabian et al. 1998) at 
41 arcmin, i.e. inside the PDS f.o.v. ($\sim1.3^{\circ}$ 
FWHM) but outside the imaging instruments' f.o.v. ($\sim28^{\prime}$ radius,
for the MECS). 
However, for a lucky coincidence, {\it Beppo}SAX observed 
both sources only four days apart: thanks to 
the collaboration between the two proposing groups, we have been
able to estimate the inter--contamination.
The strategy has been to add the LECS$+$MECS best fit model of one source
into the PDS model for the LECS$+$MECS$+$PDS fit of the other source, 
after accounting for the instrument off--axis response, and cross--checking 
the results for consistency. 
As shown in Fig. 1 (left) the PDS points remain slightly above the 
flat ($\alpha_x=0.92$) powerlaw, with no sign of declining, up to 100 keV.
This result remains unchanged even in the hypothesis, for GB 1428+422, of 
a variability of a factor $\sim$4 in 4 days, from 
$F_{1keV}=0.30\;\mu$Jy to $F_{1keV}=1.44\;\mu$Jy 
(more details in Costamante et al., in preparation).
Note that the observed X--ray flux of 1ES 1426+428 (the source is monitored 
by XTE) was not particularly high:
this suggests that the object, unlike the others ``over 100 keV" 
sources, was not in a flaring state during the {\it Beppo}SAX observation. 
It is then likely that 1ES 1426+428 is the first source found in
a ``quiescent extreme" state, offering us the best opportunity
to study the acceleration mechanism at its limit.
It is interesting also to study the relation between the broad spectral index
$\alpha_{rx}$ and the frequency of the synchrotron peak: 
compared to the HBLs data in Wolter et al. (1998), the ``extreme" BLLacs
parameters suggest a flattening at high frequencies of the correlation 
between $\alpha_{rx}$ and $\nu_{peak}$ (see Fig. 1, right). 
This behaviour  is expected in a scenario
in which the synchrotron peak moves smoothly from lower to higher energies: 
as long as the radio and X--ray bands are observing different branches of the 
synchrotron emission (before and after the peak), radio and X fluxes
change differently as the peak shifts towards higher frequencies,
thus changing $\alpha_{rx}$.
When the peak moves into the X--ray band and beyond, both bands start observing
the same branch of the emission, and so both fluxes change similarly as the peak
moves further. Consequently  $\alpha_{rx}$ stabilizes at a common (flat) value.

For all these 4 sources, the high values of the synchrotron peak frequencies
make them good candidates for TeV emission through 
the Inverse Compton mechanism.


\begin{figure}
\epsfxsize=6cm 
\vspace{-2.4cm}
\begin{tabular}{rr}
\epsfbox{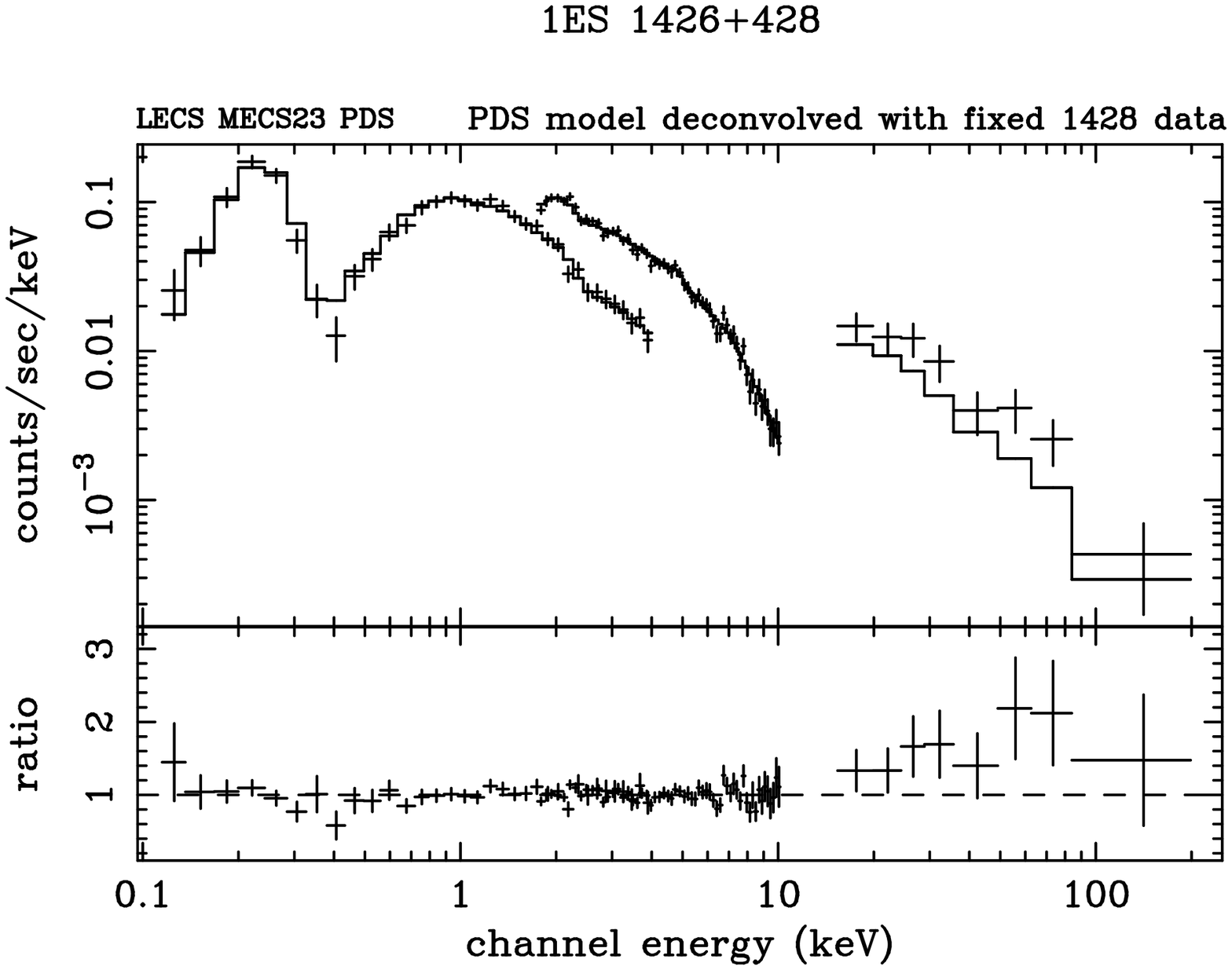}  & 
\epsfbox{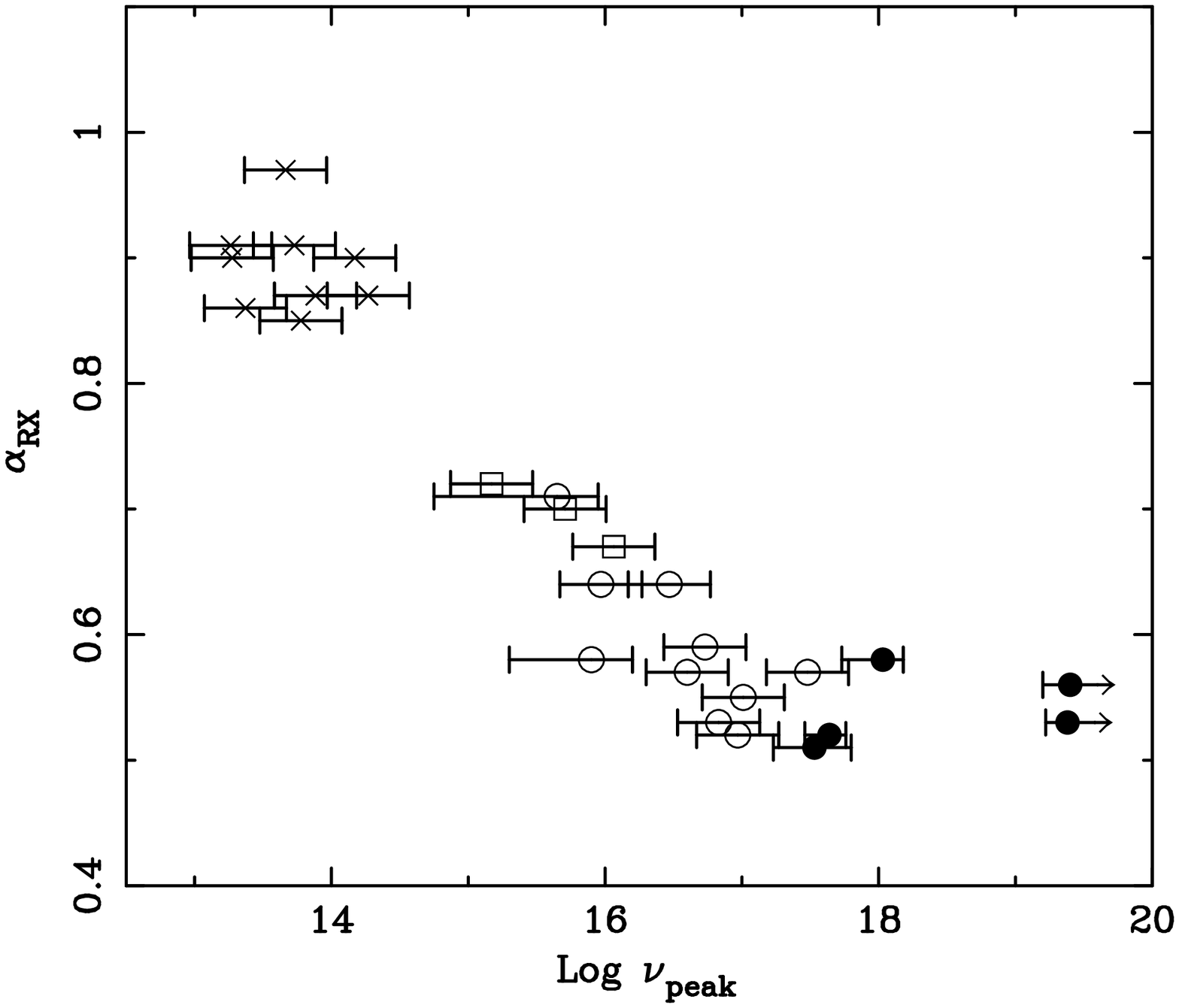} \\     
\end{tabular}
\vspace{-1.7cm}
\caption[h]{Left: LECS+MECS fit to 1ES 1426+428 X--ray data, 
 accounting for the contribution (above 15 keV) of GB 1428+422.
Right: $\alpha_{rx}$ as a function of $\nu_{peak}$. Crosses (LBL)  and 
empty squares (HBL)
are from Comastri et al. (1995); open circles (HBL) from Wolter et al. (1998). 
Filled circles are our sources and Mkn 501, in the flaring state. }
\end{figure}


\acknowledgements
L.C. thanks the STScI Visitor Program and the Cariplo Foundation for support.


\end{document}